# Ultrafast optical generation of coherent phonons in CdTe$_{1-x}$Se$_x$ quantum dots

A. V. Bragas,[1,*] C. Aku-Leh,[1] S. Costantino,[1,2] Alka Ingale,[1,3] J. Zhao,[1] and R. Merlin [1]

[1] The Harrison M. Randall Laboratory of Physics, The University of Michigan, Ann Arbor, MI 48109-1120.

[2] Departamento de Física, Facultad de Ciencias Exactas y Naturales, Universidad de Buenos Aires, 1428 Buenos Aires, Argentina.

[3] Laser Physics Division, Centre for Advanced Technology, Indore 452013, India.

TE: 734-763-4340, FAX: 734-764-5153, e-mail: abragas@umich.edu

**Abstract:** We report on the impulsive generation of coherent optical phonons in CdTe$_{0.68}$Se$_{0.32}$ nanocrystallites embedded in a glass matrix. Pump probe experiments using femtosecond laser pulses were performed by tuning the laser central energy to resonate with the absorption edge of the nanocrystals. We identify two longitudinal optical phonons, one longitudinal acoustic phonon and a fourth mode of a mixed longitudinal-transverse nature. The amplitude of the optical phonons as a function of the laser central energy exhibits a resonance that is well described by a model based on impulsive stimulated Raman scattering. The phases of the coherent phonons reveal coupling between different modes. At low power density excitations, the frequency of the optical coherent phonons deviates from values obtained from spontaneous Raman scattering. This behavior is ascribed to the presence of electronic impurity states which modify the nanocrystal dielectric function and, thereby, the frequency of the infrared-active phonons.

1. INTRODUCTION

Since the early 80s the electronic, optical and magnetic properties of systems of reduced dimensionality have been extensively studied, particularly those of nanocrystals or quantum dots (QDs). A vast literature covers topics extending from the basic physical properties of QDs to a wide range of applications (see, for example, Ref. 1-4). Due to potential applications in optoelectronic devices, doped semiconductor glasses have attracted considerable attention. Since the glass matrix is transparent in the visible range, the linear and nonlinear optical properties of these composite materials are determined by those of the semiconductor nanocrystallites. Because the emission of phonons is one of the most important electronic dephasing mechanisms, there has been much interest in the properties of phonons in QDs. The electron-phonon interaction in QDs determines the homogeneous width of the discrete electronic transitions, which, in turn, defines the oscillator strength and the performance of the QDs for optoelectronic applications. Also, the effect of confinement on the phonon modes makes them interesting on their own. Optical phonons, [5-7] confined acoustic phonons, [5, 8-11] surface phonons [9, 12] and disorder activated-phonons [13, 14] in semiconductor nanocrystallites have been extensively investigated with spontaneous Raman scattering (RS). The study of coherent phonons in semiconductor QDs, driven by ultrafast pulsed lasers, has recently begun. Coherent techniques allow the direct observation of vibrational modes in the time domain. This offers the possibility of studying phonon ultrafast dynamics as well as coherently controlling the vibrational modes. Moreover, the excitation of coherent phonons in QDs near electronic resonances allows access to the vibrational excited states, as it is well known from the extensive literature on molecular vibrons. [15, 16] There are rather few reports on coherent phonon generation in QDs. These include the generation of coherent optical phonons in nanocrystals of CdSe [17] and InP, [18] and coherent acoustic phonons in PbS, [19] PbTe, [20] PbSe, [21] and InAs [22] QDs.

## 2. COHERENT PHONON GENERATION AND DETECTION

The question of the mechanisms responsible for the generation of coherent phonons with ultrafast laser pulses has been extensively discussed in the literature in recent years. [23-29] While it is generally agreed that SRS is the driving force for the coherent phonon field $Q(\vec{r},t)$ in the spectral region where materials are transparent, there is still a controversy as to the underlying mechanism (or mechanisms) in the opaque regime. Alternative non-Raman mechanisms have been proposed involving various processes which operate on a time scale faster than the phonon period. [30] Although most of the reported resonant experiments in absorbing materials clearly support the predictions of SRS, [27, 28, 31, 32] the apparent absence of non-fully symmetric phonon modes in early experiments [33] motivated the development of an alternative model called displacive excitation of coherent phonons. [24, 25] Recently, it has been demonstrated that SRS is determined not by one but by two different tensors, one of which is the standard Raman susceptibility tensor, $\chi_{kl}^R$ (the same one that gives the cross section for spontaneous RS), and a second one, $\pi_{kl}^R$, associated with the driving force of $Q$. [32] The real parts of these two tensors are identical and, consequently, there is a single tensor for transparent materials. In this case, the light-induced driving force, proportional to $\chi_{kl}^R \equiv \pi_{kl}^R$, behaves impulsively for laser pulses that are shorter than the phonon period so that $Q \propto \sin(\Omega t)$ ($\Omega$ is the phonon frequency). On the other hand, in absorbing regions, the tensor $\pi_{kl}^R$ is the one that participates in the generation of the phonon field. The imaginary components of the two tensors in this region differ appreciably. By using the proper tensor, it is possible to reproduce the impulsive behavior in the transparent and the displacive behavior in the absorbing regime.

## A. Generation

The stimulated Raman tensor $\pi^R$ at frequency $\omega$, can be expressed approximately in terms of the dielectric function $\varepsilon(\omega)$ as [32]

$$\pi^R(\omega+\Omega,\omega) \approx \frac{\Xi}{4\pi\hbar}\left[\frac{d\,\text{Re}(\varepsilon)}{d\omega} + 2i\frac{\text{Im}(\varepsilon)}{\Omega}\right] \quad (1)$$

where $\Xi$ is the electron-phonon coupling constant (for clarity we omit the subindices). Let the sample be defined by $0 \leq z \leq l$. Then, the laser electric field for $z>0$ is given by

$$E(z,\omega) = \frac{2E(\omega)}{\eta(\omega)}\exp(-i\omega n(\omega)z/c)\exp(-\tfrac{1}{2}\alpha(\omega)z) \quad (2)$$

where $\alpha$ is the absorption coefficient. The factor $2/\eta$, where $\eta=n+i\kappa+1$, accounts for the transmission of the pulse at the front surface of the sample, $z=0$, and $n(\omega)$ and $\kappa(\omega)$ are the real and imaginary parts of the index of refraction. The phase term in Eq. (2) reflects the dependence of the field on the perturbation at the retarded time $t-nz/c$ ($c$ is the speed of light). Solving the equation of motion ignoring phonon decay [32, 34] we get

$$Q(z,t) = \frac{1}{2i}\left\{F_+(z,\Omega)\, e^{-i\Omega t} - F_-(z,\Omega)\, e^{i\Omega t}\right\} \qquad (3)$$

where

$$F_\pm(z,\Omega) = \frac{2}{\Omega}\int_{-\infty}^{\infty} d\omega \, \frac{E_0(\omega)\, E_0^*(\omega\pm\Omega)}{\eta(\omega)\,\eta(\omega\pm\Omega)} \, \pi^R(\omega,\omega\pm\Omega)\, e^{-1/2[\alpha(\omega)+\alpha(\omega\pm\Omega)]z}. \qquad (4)$$

$E_0$ is the amplitude of the electric field of the pump pulse. Notice that Eq. (4) is slightly different from the expressions shown in, [32] because we consider absorption. From (4), it follows that the components of the electric field at $\omega$ and $\omega-\Omega$, contained in the pulse spectrum, mix together to drive the coherent phonon field. The larger the spectral width, the larger the density of pairs that contributes to the generation process, resulting in a larger amplitude for $Q$. The functional form of $\pi^R$ determines $F_\pm(z,\Omega)$ which, in turn, defines the phase of $Q$. For the impulsive case, *i.e*, for below-gap excitation, $F_\pm$ is real resulting in $Q \propto \sin(\Omega t)$, whereas for the displacive case, $F_\pm$ is purely imaginary and $Q \propto \cos(\Omega t)$. In general, in the presence of damping

$$Q \propto \begin{cases} e^{-\Gamma t}\sin(\Omega t) & F_\pm \text{ real} \qquad (5a) \\ e^{-\Gamma t}\sin(\Omega t + \pi/2) & F_\pm \text{ pure imaginary} \qquad (5b) \\ e^{-\Gamma t}\sin(\Omega t + \theta) & F_\pm \text{ complex} \qquad (5c) \end{cases}$$

where the phase $\theta$ can take any value and varies strongly in regions where Im($\varepsilon$) dominates.

B. Detection

To measure the phonon field created by the pump, a second pulse, the probe, is sent at a variable time delay to interact with the phonon field. To calculate the scattered probe field we consider the nonlinear polarization proportional to the product $\chi^R Q$ and solve Maxwell equations. The complete expression for the scattered field $e_s$ at the sample exit, $z=l$, integrated over the whole length of the sample is [34]

$$e_s(\delta,\omega) = \frac{4\pi[\eta(\omega)-1]}{c\eta^2(\omega)} e^{-i\omega(n(\omega)l/c+\delta)} \left\{ \begin{array}{l} \dfrac{e_p(z=0,\omega+\Omega)}{\eta(\omega+\Omega)} \chi^R(\omega+\Omega,\omega)\, \Re_{Q_+}(l,\omega,\Omega)\, e^{-i\Omega\delta} \\ -\dfrac{e_{p0}(z=0,\omega-\Omega)}{\eta(\omega-\Omega)} \chi^R(\omega-\Omega,\omega)\, \Re_{Q_-}(l,\omega,\Omega)\, e^{i\Omega\delta} \end{array} \right\}$$

(6)

where $e_p$ is the amplitude of the probe electric field, $\delta$ is the pump-probe optical delay time,

$$\Re_{Q\pm} = \int_0^l dz\, F_\pm(z,\Omega)\, e^{-1/2[\alpha(\omega\pm\Omega)z+\alpha(\omega)(l-z)]} \quad (7)$$

and

$$\chi^R(\omega+\Omega,\omega) \approx \frac{\Xi}{4\pi\hbar}\left[\frac{d\,\mathrm{Re}(\varepsilon)}{d\omega} + i\frac{d\,\mathrm{Im}(\varepsilon)}{d\omega}\right] \quad (8)$$

is an approximate expression for the spontaneous Raman tensor. [32] From (4), (6) and (7), it follows that the probe field depends on both tensors $\chi^R$ and $\pi^R$; $\chi^R$ is associated with the detection and $\pi^R$ with the generation process.

Using (6) the spectrally integrated differential transmission, $\Delta T$ is

$$\Delta T(\delta) = \frac{c}{8\pi} \int_{-\infty}^{\infty} d\omega\, n(\omega)\left\{e_s(\delta,\omega)\, e_s^*(\delta,\omega) + 2\,\mathrm{Re}\!\left(e_s(\delta,\omega)\, e_{pT}^*(\omega)\right)\right\} \quad (9)$$

where $e_{pT}$ is the unperturbed probe field transmitted through the sample. This expression can be compared directly with the measurements.

## C. Resonant behavior: dielectric function of the composite

Our sample is a composite semiconductor-glass. Optically, we have semiconductor inclusions of dielectric function $\varepsilon_s$ in an insulating matrix of dielectric function $\varepsilon_m$, in an amount defined by the volume filing factor $f$. Eq. (1) and (7) show explicitly that the phonon field can exhibit resonant features determined by the behavior of the dielectric function $\varepsilon(\omega)$. In this work we use resonant excitation, both, to attain large phonon amplitudes and to probe the mechanism of generation. Since dielectric data for the material studied in this report is not available, we model $\varepsilon(\omega)$ following [35] and extract the dielectric function by fitting the experimental transmission of the sample. To calculate $\varepsilon(\omega)$ we assume that the crystallites are spherical and that $f$ is small enough to disregard inter-crystallite interactions. Therefore, the only disorder in our calculated $\varepsilon(\omega)$ is associated with the size dispersion in the crystallite radius. Under these conditions, the effective dielectric function of the composite is $\varepsilon(\omega) \approx \varepsilon_m [1 + 3f \int da\, \zeta(\omega,a) F(a)]$ [35], where $\zeta(\omega,a)$ is the polarizability of a sphere of radius $a$ and dielectric function $\varepsilon_s(\omega,a)$ embedded in a medium of dielectric function $\varepsilon_m$. In the calculations we use the Clausius-Mossotti formula

$$\zeta(\omega,a) = \frac{\varepsilon_s(\omega,a) - \varepsilon_m}{\varepsilon_s(\omega,a) + 2\varepsilon_m} a^3,$$

and assume that the size distribution $F$ is a Gaussian function with standard deviation $\Delta a$ and average radius $a_0$.

We assume that the laser excites only the first excitonic transition. This assumption is reasonable because the second transition is ~ 200 meV [35] above the first one. To compute $\varepsilon_s(\omega,a)$, we consider the electrons and holes confined by a spherical infinite potential and we calculate the exciton levels in the dots using the envelope and effective mass approximations. In the expression for the total energy we include a correlation correction term [35] of 65meV, which corresponds to a bulk exciton radius of ~4nm.

### D. Mode coupling

Mode coupling theory is usually invoked in situations where the Raman lineshapes show marked asymmetries or interference effects. [36-38] In particular, if a single mode couples to a continuum, the lineshape takes the form of a Fano profile, named after Fano's work on interferences of electronic transitions, [39] although the essential features of the coupling were recognized by Szigeti in his work of phonons, [40] which preceded the work of Fano. The spectral density function can be expressed in terms of the imaginary part of certain Green function. For two uncoupled modes, the relevant Green function is simply the sum of two independent contributions and, thus, the spontaneous Raman lineshape is the sum of two lorentzians. If the modes are coupled, however, the spectrum is described by a linear combination of diagonal and off diagonal terms, that depend on a set of phenomenological coupling parameters. [41] The imaginary part of this combination reproduces the asymmetric spectrum profiles. While coupled modes in the frequency domain have been understood for a long time, there is almost no discussion in the literature as to the manifestation of mode coupling in time-domain experiments[42].

Consider two coupled oscillators, $Q_1$ and $Q_2$, of frequency and damping $\Omega_1, \Gamma_1$ and $\Omega_2, \Gamma_2$ respectively. For impulsive excitation in the domain of transparency, the most general solution for $t>0$ is of the form

$$Q_1 = A_{11} \exp(-\Gamma_1' t) \sin(\Omega_1' t + \varphi_{11}) + A_{12} \exp(-\Gamma_2' t) \sin(\Omega_2' t + \varphi_{12})$$
$$Q_2 = A_{21} \exp(-\Gamma_1' t) \sin(\Omega_1' t + \varphi_{21}) + A_{22} \exp(-\Gamma_2' t) \sin(\Omega_2' t + \varphi_{22})$$

(10)

where the renormalized frequencies $\Omega_j'$ and dampings $\Gamma_j'$, as well as the ratios $\frac{A_{11}}{A_{21}}$ and $\frac{A_{12}}{A_{22}}$ and the phase differences $\varphi_{11} - \varphi_{21}$, $\varphi_{12} - \varphi_{22}$ are all functions of the coupling parameters. [41] The signal of interest in time-domain experiments is a linear combination of $Q_1$ and $Q_2$, of the form

$$\sum_k B_k \exp(-\Gamma_k' t) \sin(\Omega_k' t + \phi_k) \qquad (11)$$

where $k=1,2$ label the two eigenmodes. For independent oscillators $A_{12} = A_{21} = 0$, and the boundary conditions $Q_1=Q_2=0$ at $t=0$ give $\varphi_{11} = \varphi_{22} = 0$ and, therefore, $\phi_1 = \phi_2 = 0$. Hence, we recover a simple sum of two oscillators of the form shown in Eq. (5a). Some reflection shows that $\phi_1 = \phi_2 = 0$ also applies to coupled modes in the absence of dissipation, since the coupling leads only to a renormalization of the frequencies. However, the combination of coupling and dissipation invariably leads to $\phi_k \neq 0$. We see from the comparison of Eq.(5a) and Eq.(11) that the unique signature of the coupling is that the initial phases of the oscillations are not equal to zero. These consideration apply to transparent substances. In the general case of complex $F_\pm$ [Eq. (5c)], coupling introduces an additional phase to the phase predicted by SRS theory for independent oscillators.

## 3. EXPERIMENTAL

Our sample is a commercial 3mm-thick RG780 filter (Schott Glass Technologies, Inc.) made of semiconductor nanocrystallites of $CdTe_{0.68}Se_{0.32}$ embedded in a borosilicate matrix. [6] Bulk $CdTe_{1-x}Se_x$ crystallizes in the zincblende structure for $x<0.36$. [44, 45] The energy bandgap $E_g$ versus composition displays a bowing effect [46] and, at $x=0.32$, $E_g=1.4$eV. The electronic and hole spatial confinement in the dots produces a blue-shift of the gap, which can be determined by measuring the sample transmission, as shown in Fig. 1.

Using the parameters shown in Table I, we computed the dielectric function of the composite $\varepsilon(\omega)$ following the model of Section 2C. From Im($\varepsilon$) we fit the experimental data. The best fit parameters are listed in Table II. The inset of Fig. 1 shows the calculated real and imaginary parts of $\varepsilon(\omega)$. The fit involves four parameters: $\Delta a$, $f$, $\gamma$, the electronic line-width, and $E_1$, the first exciton energy. The values obtained are consistent with those reported for similar materials. [7, 47, 48, 49]

Ternary alloys exhibit either one or two-mode phonon behavior.[50] $CdTe_{1-x}Se_x$ shows two-mode behavior leading to two, each, CdSe-like and CdTe-like longitudinal optical (LO) and transverse optical (TO) phonons. The frequencies of these phonons vary with the alloy composition. For $x$=0.32, the frequencies are 190cm$^{-1}$(LO) and 170cm$^{-1}$(TO) for the CdSe-like, and 160cm$^{-1}$(LO) and 140cm$^{-1}$(TO) for the CdTe-like phonons.[44] Additional modes attributed to clustering or disorder have been reported in bulk $CdTe_{0.65}Se_{0.35}$,[45] and also for related ternary compounds.[51]

In our pump-probe experiments we used a modelocked Ti:Sapphire laser (Tsunami, Spectra Physics), operating at 82MHz repetition rate, which emitted pulses of 45-50 fs, tunable between 1.55 and 1.62 eV. This range covers the region around the main absorption edge of the sample (See Fig.1). Raman measurements were recorded with a triple grating modular XY Dilor Spectrometer, and a continuous wave home-made Ti:Sapphire laser, also tunable in the same range. All the experiments were performed at room temperature.

A typical setup is shown schematically in Fig. 2. The probe intensity is six times lower than the pump intensity. The polarizations of the pump and probe beams are perpendicular to each other in order to reduce the scattering of the pump into the detector. The transmission is measured using photodiode detectors, by balancing a portion of the probe signal before the sample with the total transmission after the sample. Lock-in detection is performed by modulating the pump intensity with a mechanical chopper at 3 kHz.

A. Mode assignment

Fig. 3 shows the differential probe transmission, $\Delta T$, as a function of the pump-probe delay at the laser central energy $E_c$= 1.587eV. The magnitude of $\Delta T/T$, where $T$ is the total probe transmission, is independent of the probe intensity and gives a measure of the strength of the signal. The oscillations due to coherent phonons are superimposed on an exponential background, arising from electronic excitations. For clarity, this electronic contribution is removed in the following from the total signal to isolate the phonon oscillations.

We fit our data using linear prediction (LP) methods, [52] to expressions of the form of Eq. (11). Using this fitting procedure we can obtain the amplitude, phase, damping and frequency of an arbitrary number of different modes. In Fig. 4 we show an example of the fit and, separately, the individual contributions of the various modes. Four phonon modes can be distinguished: two LO modes at $\Omega$=197cm$^{-1}$ and $\Omega$=162 cm$^{-1}$, one confined longitudinal acoustic (LA) mode at $\Omega$=24 cm$^{-1}$ and a fourth mode at $\Omega$=135cm$^{-1}$. The behavior of the confined acoustic phonon will be discussed elsewhere. [53] While the frequency of the fourth mode is close to the CdTe-like TO phonon frequency, [44] we believe that it has a mixed TO-LO character, [54] as it is labeled in Fig. 4, because LO modes usually dominate the resonant spectra. [13, 54] A detailed discussion for this mode is given in Section 3C.

B.  Resonant behavior

In Fig. 5 we show time domain data for different central laser energies, $E_c$. The pump intensity is set at 2kW/cm$^2$. Special care was taken in keeping the pulsewidth nearly constant at about 45-50 fs. Since the transmission $T$ is very small near $E_1$, we present results for $\Delta T$, to avoid errors due to the division by $T$.

The dependence of the intensities of the CdSe-like and CdTe-like LO modes on $E_c$ is shown in Fig. 6, using the values of the amplitudes in Eq. (11) from LP fitting. The calculated SRS signals from (9) are shown a full lines. The measured differential transmission exhibits a shift on the order of the LO frequency with respect to the calculations. Such an effect has been reported in spontaneous resonant Raman experiments [32] and attributed to vertex corrections. For that reason, in Figure 6, the calculated curve has been shifted to lower energies by 20 meV, in order to match the experimental curve. Other than the fact that the experimental width of the resonance is slightly smaller than the calculated one, the agreement between theory and experiment is quite good. The ratio of the maxima in Fig. 6 gives an experimental value for the ratio of the Fröhlich polaron constants $\alpha_{CdSe}$ and $\alpha_{CdTe}$ for CdSe and CdTe, respectively. We find

$$\frac{\alpha_{CdSe}}{\alpha_{CdTe}} \approx \sqrt{\frac{\Delta T_{CdSe}}{\Delta T_{CdTe}} \frac{\Omega_{CdSe}}{\Omega_{CdTe}}} = 1.8$$, which is within 30% of the value gained from Ref. 46.

The calculated phases for the two LO modes, shifted as in Fig.6, are shown as solid and dotted lines in Fig. 7. We assign an error to the phase equal to the acquisition step of the pump-probe delay multiplied by the phonon frequency. As it can be seen, the agreement between experiment and theory is poor. There are a number of factors which make the determination of the phase difficult. The main source of error is related to the determination of *time zero*, i.e. the point at which pump and probe overlap in time, which fixes the value of the initial phase. In our experiments, an error of 12 fs in the delay gives a 15$^o$ error in the phase. Usually, time zero is taken at the position of the peak of the crosscorrelation measured in a non-linear crystal. In non-collinear pump-probe geometries, the differences in the refractive index of the sample and the non linear crystal lead to different internal angles. In turn, this causes differences in the position at which the pulses overlap that can shift time zero (3.7 μm is equivalent to ~15$^o$). Moreover, the definition of time zero as the maximum of the crosscorrelation curve is somehow arbitrary. All this can introduce additional systematic errors, that shift the curves rigidly up or down. Another source of error is related to the fact that the electronic background under resonant conditions is much larger than the phonon signal. After a few tens of femtoseconds, the electronic response of the material is well described by a multi-exponential decay. However, there is no reason to believe that such a fit can be extended down to the first tens of femtoseconds after time zero. This can distort the shape of the phonon signal after subtraction of the electronic contribution, introducing errors in the fit parameters, among which the initial phase is the most sensitive. Nevertheless, systematic errors cannot explain the gap between the phases of the two LO modes in Fig. 7. As discussed in Section 2D, coupling between modes affects the phase of the oscillators, and we believe that the observed phase difference is the manifestation in time domain of the coupling between the CdTe-like and TO-LO modes. This issue is discussed in the next subsection.

Fig. 8 shows the dependence of the frequencies of the optical modes with the laser central energy $E_c$. In Fig. 8 (a) the CdSe-like LO phonon frequency exhibits a small but measurable red-shift. For smaller dots, the frequency is expected to be smaller than for larger dots, due to phonon confinement and the negative dispersion of the LO phonons. The observed red-shift is consistent with the fact that smaller dots become resonantly excited when the laser energy increases. Changes in absorption along the absorption edge can also contribute to changes in frequency of the same sign as the confinement. In Figs. 8 (b) and (c), we plot the frequencies of the other two optical modes as a function of $E_c$. The CdTe-like LO phonon frequency is almost independent of $E_c$ but the TO-LO mode exhibits a large blue-shift which is inconsistent with the previous analysis for the CdSe-like mode. As discussed later in Section 3C, we attribute such a behavior to effects due to coupling.

Because $Q$ depends on the QD size (and also the shape), the transmitted signal is a weighted average of contributions of excited nanocrystallites inside the composite. The permittivity and, hence, the Raman susceptibility tensors of an individual dot have specific resonant signatures. In Fig. 9 we show the calculated $\chi^R$ and $\pi^R$ for a QD with radius equal to $a_0$ (3.9nm). Notice that the position of the resonance depends on the radius. Fig. 10 shows the calculated $\Delta T$ as a function of probe delay considering size dispersion (but ignoring the optical mode dispersion) for laser central energies ranging from 1.55 eV to 1.68 eV. The laser spectrum is modeled as a Gaussian function with bandwidth of 20nm. The amplitude of $\Delta T$ shows a resonant behavior, which is mostly dominated by the frequency derivative of the total transmission of the sample, as can be seen from (6) and (7). This is due to the fact that the spectrum of the scattered field $e_s$ oscillates as a function of the delay, with amplitude proportional to $\Omega$. As the sample itself acts as a filter, the detector recording the integrated transmission measures these oscillations as changes in amplitude, which are largest when the derivative of the transmission is maximum. The phase behavior is non trivial near the absorption edge where terms from different dots add to the total phase. We notice that, whereas $Q \propto \sin(\Omega t)$, $\Delta T$ in *transmission experiments* behaves as a cosine-like function in the transparent region (~1.55eV). [23] It can also be shown that in the absorbing region $\Delta T$ is a sine-like function (~1.68eV). This is because the scattered probe field $e_s$, Eq. (6), is obtained from a wave equation containing the time-derivative of $Q$.

### C. TO-LO mode and mode coupling

The frequency of the mode labeled TO-LO in Fig. 4 increases from ~130cm$^{-1}$ to ~140 cm$^{-1}$ when the laser central energy increases from 1.55eV to 1.61eV (Fig. 8 (c)), while the line width also increases from ~10cm$^{-1}$ to ~17cm$^{-1}$. This mode is not detectable above the noise level when the power density is below ~500W/cm$^2$. As shown in Fig. 4, the salient feature of this mode in pump-probe measurements is that it vibrates out of phase with respect to the LO modes.

We assign the feature at ~140cm$^{-1}$ to an $\ell \geq 1$ optical confined mode where $\ell$ is the pseudo angular momentum. [54] Since the frequency of this peak is between those of the TO and LO CdTe modes (and much closer to the TO), we believe that this mode, we label TO-LO, has a mixed character with transverse, longitudinal and surface contributions. [54] Our experiments cannot entirely rule out the possibility that the ~140cm$^{-1}$ feature is a disorder activated mode even though there is no evidence of such modes in bulk materials (a mode due to clustering of the anions around the cations has been previously observed in bulk CdTe$_{0.65}$Se$_{0.35}$ [45] but at ~ 180 cm$^{-1}$). However, we note that disorder associated with size or imperfect crystalline order could give rise to additional peaks in the Raman spectrum of QDs, as studied in Ref. 14.

Irrespective of the origin of the 140cm$^{-1}$ mode, we can account for some of its features. The initial phase shift of this mode can be explained as due to coupling to another mode, specifically, the CdTe-like LO phonon since, in the absence of coupling, all optical modes are expected to show the same initial phase. The unusual behavior shown in Fig. 8(c) is also consistent with our interpretation in that the blue-shift may be accounted for by assuming a radius-dependent coupling. Spontaneous Raman measurements add more evidence to the coupling interpretation. The Raman spectrum in Fig. 11 shows a low-frequency asymmetry in the CdTe-like phonon profile, and a weak dip at the position of the TO-LO mode. The curves in the inset are calculated Raman spectra. The top spectrum is for three independent modes whereas the bottom spectrum shows a complex lineshape which results from introducing coupling between the two lowest frequency modes. The measured spectrum is quite similar to the Raman profile for coupled modes. Other than first-order scattering, we observe several second order features above ~250 cm$^{-1}$ and, in particular, the first overtone of the TO-LO mode at ~270cm$^{-1}$. The fact that this feature is as strong as the first order one and narrower, reinforces the idea that the TO-LO and the Cd-Te like LO phonon are coupled. We note that neither Raman nor pump-probe data shows strong indication of the existence of a TO-LO mode associated with CdSe although the Raman spectrum shows a weak low-frequency asymmetry for the CdSe-like LO phonon. It is not clear whether this feature is due to the disorder mode seen in bulk CdTe$_{0.65}$Se$_{0.35}$[45] and in QDs of similar ternary alloys,[13] or to confined modes.[54]

D. Power dependence of LO modes

The laser power dependence of the LO modes was studied for power densities in the range 10 to 10kW/cm². The amplitude of the modes varies linearly with power density up to 2kW/cm². Fig. 12 shows the frequency of the CdSe-like LO mode as a function of power density for both pump-probe and spontaneous Raman measurements. The most significant feature of these results is the marked discrepancy between Raman and pump-probe frequencies at low excitation powers. The slight but systematic difference of less than 0.5cm⁻¹ at higher densities can be accounted for by the fact that pump-probe experiments tests ground as well as excited states near an electronic resonance, while spontaneous Raman only probes ground states, [15, 17] and the decrease of the Raman frequency with increasing power density is attributed to laser heating (due to power limitations of the pulsed laser, we did not study the high power regime in the time domain). At low power densities, the CdSe-like LO frequency obtained from pump-probe measurements changes by about 2%. The CdTe-like LO frequency also changes at approximately the same rate.

We ascribe the differences at low power densities to the existence of defects or impurities, possibly located at the surface of the QD. [55] We model the defect as a two-level system and we assume that the energy difference between the levels is on the order of the LO energy. Fig. 13 shows the energy level diagram we use for our calculations. An electron initially in the ground state $|0\rangle$ of the QD is photoexcited to $|1\rangle$ and becomes trapped by the impurity in level $|0_d\rangle$ at the rate $\gamma_t$. Finally, it relaxes to $|0\rangle$ at the rate $\gamma_r$. In the presence of trapped electrons, the dielectric function of the semiconductor in the infrared is modified as:

$$\varepsilon_{IR}(\omega) = \varepsilon_\infty \left(1 + \frac{\omega_{LO}^2 - \omega_{TO}^2}{\omega_{TO}^2 - \omega^2 + i\omega\Gamma}\right) + \frac{n_0 S \omega_d^2}{\omega_d^2 - \omega^2 - i\omega\Gamma_d} \quad (12)$$

Here $\omega_{LO}$ and $\omega_{TO}$ are the intrinsic values of the LO and TO frequencies, $\omega_d$ is the impurity frequency, $\Gamma$ and $\Gamma_d$ are, respectively, the phonon and impurity damping, $S$ is the strength of the impurity transition and $n_0$ is the quantum mechanical probability for the trapped electron to be in the state $|0_d\rangle$. The new longitudinal modes can be obtained by setting $\varepsilon_{IR} = 0$. A simple inspection shows that if $\omega_d > \omega_{LO}$ the frequency of the renormalized mode shifts to lower values and depends on power through $n_0$.

The dynamics of the electron populations $n_1$ and $n_0$ of levels $|1\rangle$ and $|0_d\rangle$ can be described by the rate equations:

$$\frac{dn_1}{dt} = \lambda P(t) - \gamma_t n_1 \qquad (13)$$

$$\frac{dn_0}{dt} = \gamma_t n_1 - \gamma_r n_0 \qquad (14)$$

where $P$ is the laser power density and $\lambda$ is the absorption coefficient. For a cw source, the steady-state population of $n_0$ is $\frac{\lambda P}{\gamma_r}$ and, therefore, the impurity level is full if $\frac{\lambda P}{\gamma_r} \geq 1$. Our results require that $P > P_{sat} = \frac{\gamma_r}{\lambda}$ for the whole range of power densities, so that the trap is always full. Within this interpretation, the frequency measured by spontaneous RS is not $\omega_{LO}$, i.e., the bare LO mode frequency, but the frequency of the optical mode predicted by (12).

For the impulsive case $P(t) = \frac{P}{\Gamma_{rep}} \delta(t)$, where $\Gamma_{rep}$ is the laser repetition rate and $\delta(t)$ is the Dirac delta function. Hence, for $\gamma_r \ll \gamma_t$, $n_0(t) = \frac{\lambda P}{\Gamma_{rep}} \left(1 - e^{-\gamma_t t}\right)$. The power for which the population is equal to one at large times is $P_0 = \frac{\Gamma_{rep}}{\lambda}$. If $P > P_0$ for the whole range, then:

$$n_0(t) = \begin{cases} \frac{\lambda P}{\Gamma_{rep}} \left(1 - e^{-\gamma_t t}\right) & \text{if } t < t_{crit} \\ 1 & \text{if } t \geq t_{crit} \end{cases} \quad (15)$$

where $t_{crit} \cong \frac{P_0}{P} \frac{1}{\gamma_t}$ is the power dependent time at which the trap fills (see inset of Fig. 13). If $t_{crit} > \tau_p$, where $\tau_p$ is the decay time of the phonon, the intrinsic LO frequency is not modified since the phonon is not aware of the presence of the trapped electron before it decays whereas, for $t_{crit} < \tau_p$, the phonon frequency has the same value as that measured in cw experiments. The former condition is true if $P < \frac{P_0}{\gamma_t \tau_p} = \frac{\Gamma_{rep} P_{sat}}{\gamma_r \gamma_t \tau_p}$. For our experiment $\tau_p \cong 1$ ps and $\Gamma_{rep}^{-1} = 10$ ns. If we take the reasonable values $P_{sat} = 50$ W/cm$^2$, $\gamma_r^{-1} = 1$ ns and $\gamma_t^{-1} = 50$ ps,[56] the condition for obtaining the bare LO frequency in time domain is $P < 250$ W/cm$^2$, which is consistent with the behavior shown in Fig. 12.

## 4. CONCLUSIONS

Four different coherent phonons have been generated in a semiconductor doped glass by femtosecond laser pulse excitation: two longitudinal phonons, one acoustic phonon and one confined optical mode of mixed TO-LO character. Our experimental measurements support the results predicted by the SRS model in a region of laser excitation where the absorption varies and non-trivial changes of the amplitude and phase occur. The inhomogeneous distribution of the dot size, the width of the sample, and the stimulated Raman tensor were taken into account to perform the SRS calculations. The small changes in frequency of the LO modes due to confinement were not considered. The calculation gives a complete description of the behavior of the amplitude and the phase of the coherent LO phonons. The measurements for CdSe-like LO and Cd-Te LO modes show a marked resonance which agree fairly well with the theoretical predictions.

The difficulty in precisely determining the initial phase of the phonons has been discussed. Since the phase is extremely sensitive to changes in any physical interaction and experimental conditions, care must be exercised in analyzing its behavior. There is a small discrepancy between the model and the measured phase for the LO phonons, although this difference can be partly explained by the errors arising from the time zero indetermination. The deviation is unquestionably large for the phase of the TO-LO mode. This mode oscillates out of phase with the LO phonons. We ascribed the phase shift to coupling with the CdTe-like LO phonon.

Finally, the frequency of the LO measured by pump-probe experiments show deviations from the Raman frequencies at low power excitations. We assigned this behavior to the presence of electronic impurities which modify the dielectric constant of the semiconductor in the infrared.

We showed that RS measures the optical frequency modified by the presence of the impurity whereas pump-probe shows the value of the intrinsic LO frequency below certain critical power.

ACKNOWLEDGEMENTS

This work was partially supported by the AFOSR under contract F49620-00-1-0328 through the MURI program. One of us (AVB) acknowledges partial support of Consejo Nacional de Investigaciones Científicas y Técnicas, CONICET, Argentina.

**TABLE I.** Parameters used to calculate the dielectric function of the composite.

| | |
|---|---|
| CdSe gap energy, | 1.74 eV [46] |
| CdTe gap energy | 1.5 eV [46] |
| CdTe$_{0.68}$Se$_{0.32}$ gap energy, $E_g$ | 1.4 eV [46] |
| Electronic effective mass | 0.107 $m_0$ [a, b, 57] |
| Heavy hole effective mass | 0.7 $m_0$ [b, 57] |
| High-frequency dielectric constant, $\varepsilon_{s\infty}$ | 7 [c, 57] |
| Static dielectric constant, $\varepsilon_{s0}$ | 10 |
| Glass dielectric constant, $\varepsilon_m$ | 3 |

[a] $m_0$ is the free electron mass.
[b] interpolated from CdSe and CdTe values.
[c] average value.

**TABLE II.** Parameters obtained after fitting the sample transmission.

| Parameter | Value |
|---|---|
| First excitonic transition, $E_1$ | 1.68 eV |
| Average radius, $a_0$ | 39 Å |
| Electronic linewidth, $\gamma$ | 18 meV |
| Filling factor, $f$ | 0.075 |
| Size dispersion, $\Delta a$ | 10% |

FIGURE CAPTIONS

FIG. 1. Transmission of the RG780 Schott filter near the main absorption edge. The composite is made of semiconductor nanocrystals of $CdTe_{0.68}Se_{0.32}$, immersed in a glass matrix. Experimental values are indicated with filled circles. The line is a fit to the model explained in the text, taking into account the size dispersion of the dots, and with the parameters from Tables I and II. The first excitonic transition is set at 1.68eV, corresponding to an average dot size of 39Å. The inset shows the calculated real and imaginary parts of the dielectric function $\varepsilon(\omega)$ of the composite.

FIG. 2. Schematics of a pump-probe experiment. BS: beam splitter, $\lambda/2$: half waveplate. The pump pulse is modulated with a mechanical chopper. The probe pulse goes through the optical delay line. Pump and probe beam polarizations are perpendicular to each other, to reduce the pump scattering into the detector. The total transmission of the sample is measured using a balanced photodiode detector.

FIG. 3. Coherent phonon signal for $CdTe_{0.68}Se_{0.32}$ QDs at room temperature. Typical differential transmission signal $\Delta T$, taken at laser central energy $E_c$=1.587eV. The coherent phonon vibrations are superimposed on an exponential electronic background.

FIG. 4. The top trace shows the total normalized differential transmission, $\Delta T/T$, with the electronic background subtracted. The full circles are the experimental points. The superimposed line is the fit using linear prediction (LP) methods. The lower panel shows the four contributions to the total signal gained from the subtraction of the fits of the remainder modes from the total experimental signal. The modes are: two longitudinal optical, CdSe-like and CdTe-like phonons, a mixed TO-LO mode and confined acoustic phonon.

FIG. 5. Differential transmission $\Delta T$ as a function of the probe delay for different laser central energies $E_c$. The data was normalized to a common laser power. The curves are shifted vertically for clarity.

FIG. 6. Amplitude of the LO coherent phonons as a function of $E_c$. The full squares and the open circles are the amplitudes for the LO optical modes, gained from the LP fitting. Solid lines show the impulsive stimulated Raman scattering (SRS) calculation, shifted to lower energies by 20 meV. The CdSe-like LO phonon energy is indicated as $\hbar\Omega \approx 25 meV$.

FIG. 7. Phase of the experimental signal for the LO modes as a function of $E_c$. The solid (dashed) line is the SRS phase calculation for the CdSe-like (CdTe-like) LO.

FIG. 8. Frequency of the LO phonons as a function of $E_c$. (a) CdSe-like LO phonon, (b) CdTe-like LO phonon and (c) TO-LO mode. Full circles are experimental values and dashed lines are guides to the eye.

FIG. 9. Real and imaginary parts of the stimulated $\pi^R$ and spontaneous $\chi^R$ Raman susceptibility tensors. The calculation was performed using the dielectric function of a semiconductor QD of radius 39Å (average radius of the sample). The tensors differ in their imaginary parts. Note that, in particular, at the resonance frequency, Im($\pi^R$) is maximum whereas Im($\chi^R$) is zero.

FIG. 10. Calculated total differential transmission $\Delta T$ for a single mode as a function of the probe delay for different central laser energies, $E_c$. The calculation takes into account the dot size dispersion. A single phonon frequency was considered for all the dot sizes. The arrow indicates the range of $E_c$ covered by our experiments. A clear resonant behavior is shown.

FIG. 11. Spontaneous Raman spectrum recorded at 514 nm. A weak indication of the TO-LO is seen at ~140 cm$^{-1}$ and the second order of this mode is resolved at 270cm$^{-1}$. The CdTe-like LO lineshape shows an asymmetry due to coupling with the TO-LO mode. The upper trace of the inset shows the calculated Raman lineshape for three independent oscillators. The lower trace of the inset shows an example of the Raman lineshape for two coupled oscillators and one independent mode.

FIG. 12. Power dependence of the frequency of the CdSe-like LO phonon, for Raman and pump-probe measurements. Spontaneous Raman data was taken at 1.625eV and pump-probe measurements at $E_c$=1.587eV. A correction due to the differences in absorption was included to calculate the power density. The solid line superimposed to the Raman measurements is a guide to the eye. The Raman frequency is nearly constant for low power densities, but decreases for higher densities because of local heating of the sample. Pump-probe frequency increases for low power densities.

FIG. 13. Energy diagram of the QD with a single defect. An electron in the ground state $|0\rangle$ of the QD is photoexcited to $|1\rangle$ and becomes trapped by the impurity level $|0_d\rangle$ at the rate $\gamma_t$. Eventually, it relaxes to $|0\rangle$ at the rate $\gamma_r$. The frequency $\omega_d$ is on the order of the LO frequencies. In the inset, the population of the level $|0_d\rangle$ in time domain experiments is shown as a function of time; $t_{crit}$ is the time it takes to reach the saturation value $n_0$=1, which corresponds to a power density $P_0$. The power density $P$ is always larger than $P_0$.

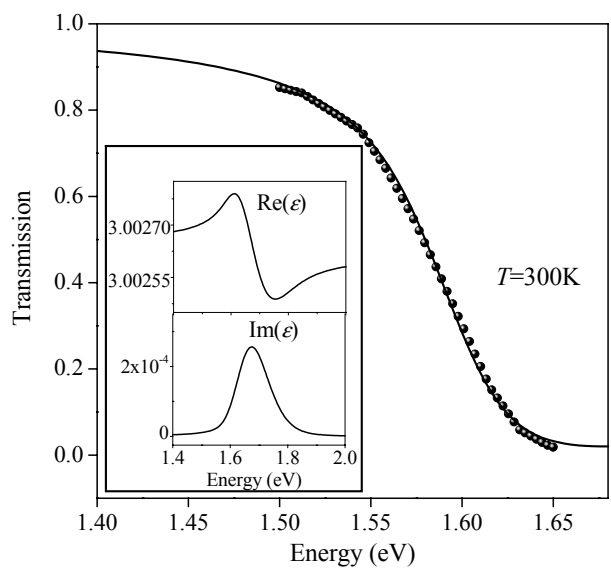

Figure 1

Bragas *et al*.

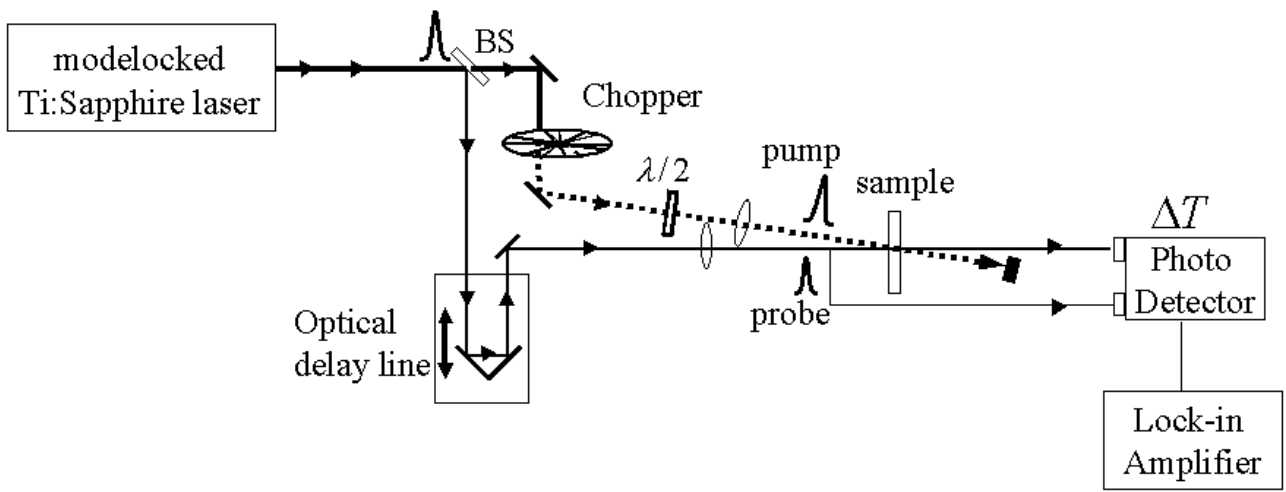

Note: two-column journal size

Figure 2

Bragas *et al*.

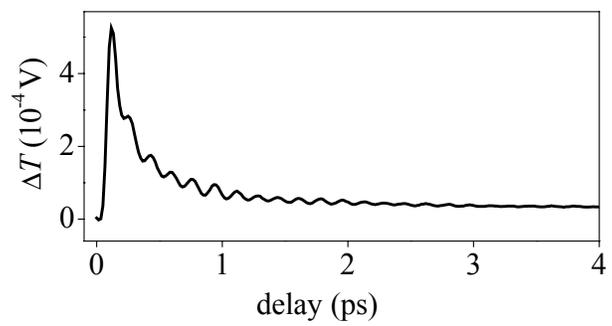

Figure 3

Bragas *et al*.

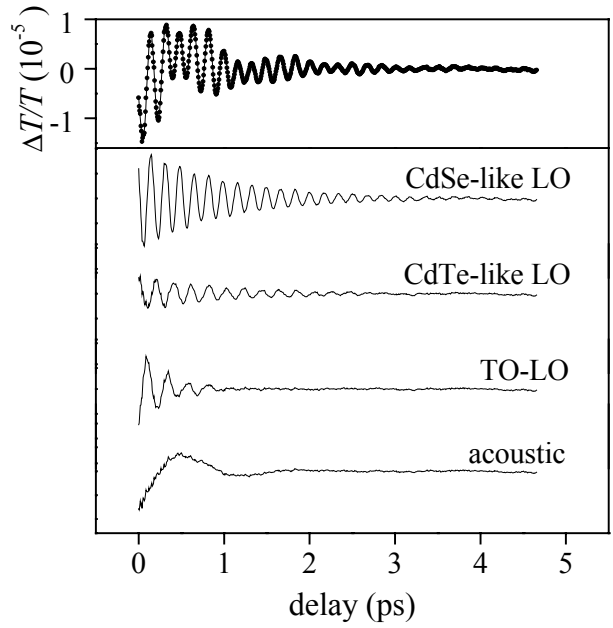

Figure 4

Bragas *et al*.

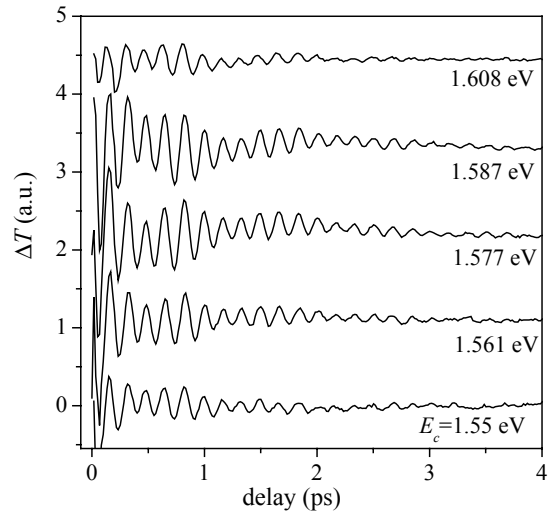

Figure 5

Bragas *et al*.

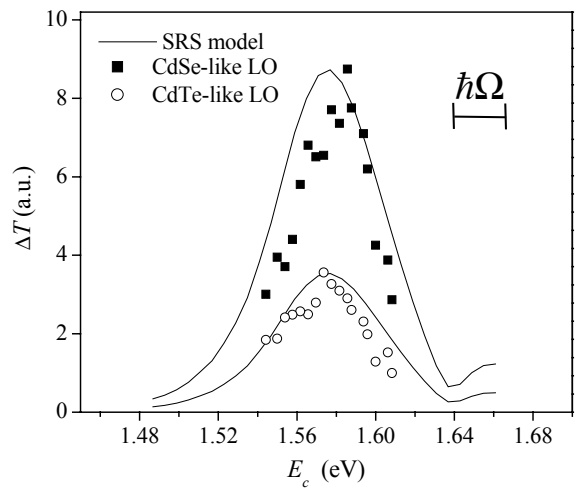

Figure 6

Bragas *et al*.

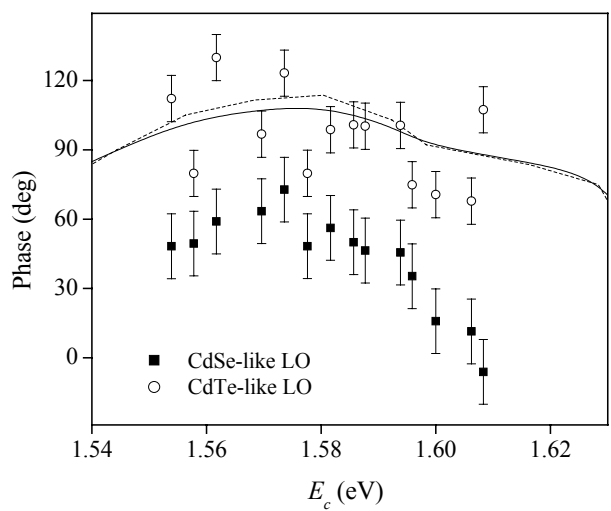

Figure 7

Bragas *et al*.

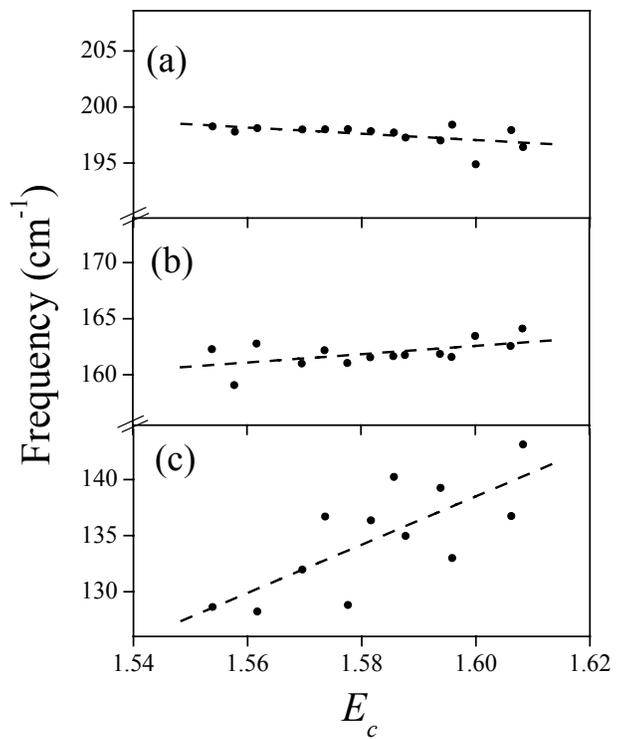

Figure 8

Bragas *et al*.

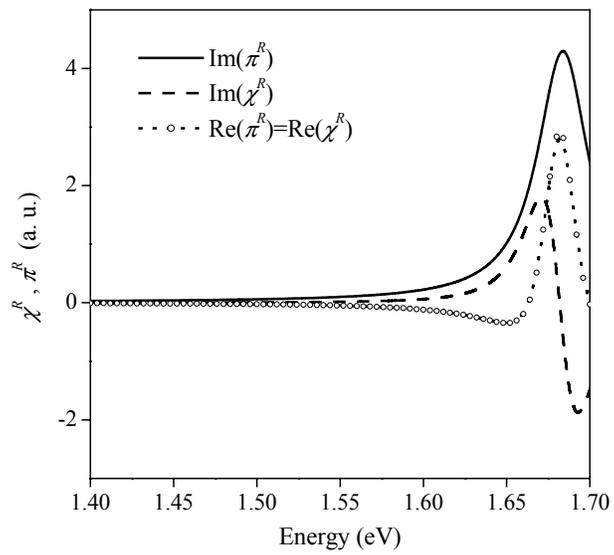

Figure 9

Bragas *et al*.

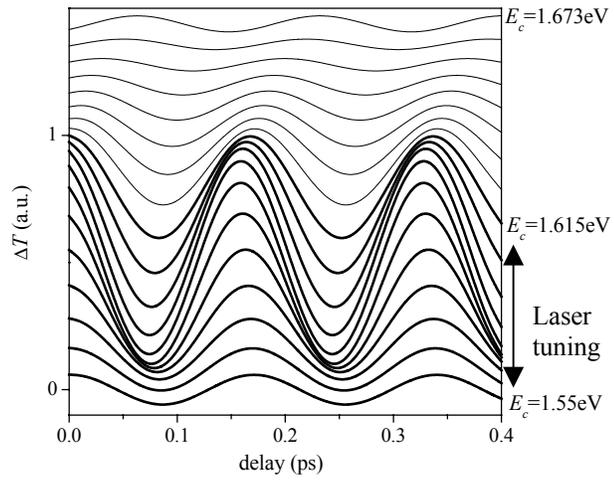

Figure 10

Bragas *et al*.

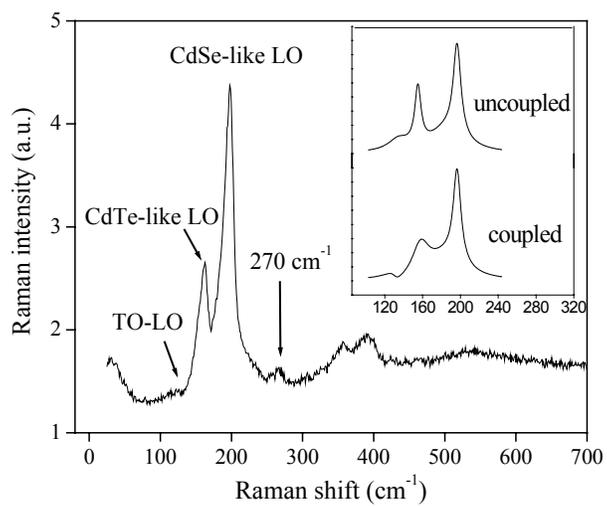

Figure 11

Bragas *et al*.

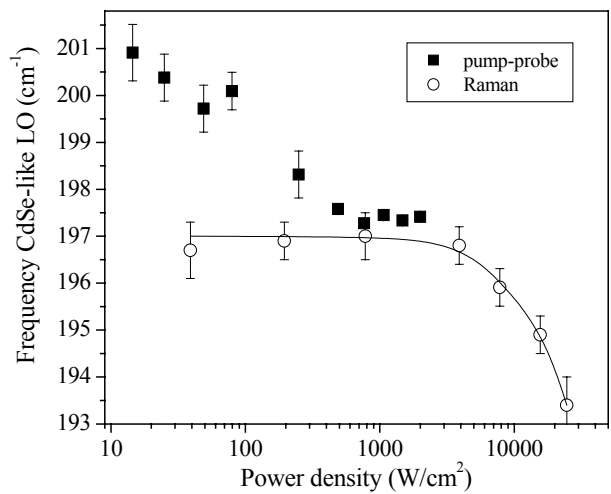

Figure 12

Bragas *et al*.

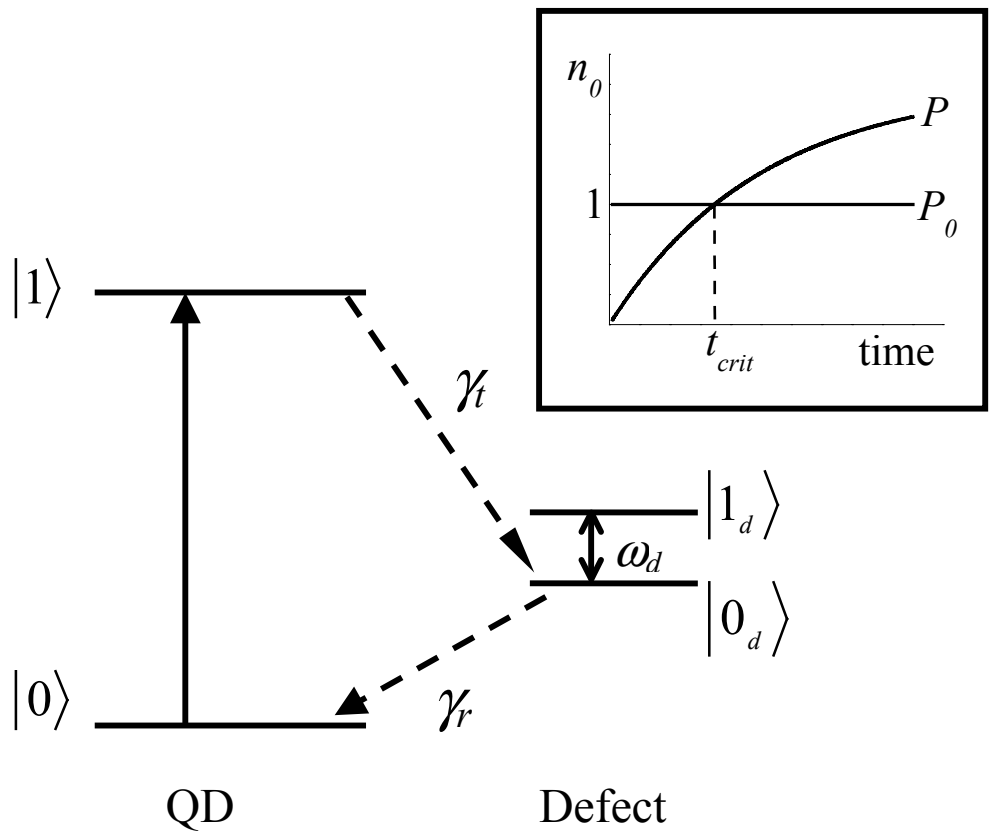

Figure 13

Bragas *et al*.